\documentclass[journal]{IEEEtran}

\IEEEoverridecommandlockouts
\usepackage{cite}
\usepackage{amsmath,amssymb,amsfonts}
\usepackage{algorithmic}
\usepackage{graphicx}
\usepackage{textcomp}
\usepackage{xcolor}
\usepackage{comment}
\def\BibTeX{{\rm B\kern-.05em{\sc i\kern-.025em b}\kern-.08em
    T\kern-.1667em\lower.7ex\hbox{E}\kern-.125emX}}

\usepackage[utf8]{inputenc}
\usepackage{xr}
\usepackage{mathtools}
\usepackage{enumitem}
\usepackage{ragged2e}
\usepackage{authblk}
\usepackage{subcaption}
\usepackage{wrapfig}
\usepackage{mathptmx} 
\usepackage{times} 
\usepackage{cite}
\usepackage{graphicx}
\usepackage{epstopdf}
\usepackage{ifpdf}
\usepackage{graphics}
\usepackage{verbatim}
\usepackage{setspace}
\usepackage{url}
\usepackage{color}
\usepackage{breqn}
\usepackage{xr}

\newtheorem{theorem}{Theorem}[section]
\newtheorem{lemma}[theorem]{Lemma}

\usepackage{multirow}

\begin{document}

\title{Towards Trust and Reputation as a Service in a Blockchain-based Decentralized Marketplace}

\author{Stephen Olariu}
\author{Ravi Mukkamalai}
\author{Meshari Aljohani}
\affil{Department of Computer Science, Old Dominion University\\ Norfolk, Virginia, 23529, USA}
\affil{\{solariu, rmukkama, maljo001\}@odu.edu}

\maketitle

\begin{abstract}
Motivated by the challenges inherent in implementing trusted services in the Society 5.0 initiative, we propose a novel trust and reputation 
service for a decentralized marketplace. We assume that a Smart Contract is associated with each transaction and that the Smart Contract 
is responsible for providing automatic feedback, replacing notoriously unreliable buyer feedback by a more objective assessment of how 
well the parties have fulfilled their obligations. Our trust and reputation service was inspired by Laplace’s Law of Succession, where 
trust in a seller is defined as the probability that she will fulfill her obligations on the next transaction. We  offer three applications. 
First, we discuss an application to a multi-segment marketplace, where a malicious seller may establish a stellar reputation by selling cheap 
items, only to use their excellent reputation to defraud buyers in a different market segment. Next, we demonstrate how our trust and 
reputation service works in the context of sellers with time-varying performance by providing two discounting schemes wherein older 
reputation scores are given less weight than more recent ones. Finally, we show how to predict trust and reputation far in the future, 
based on incomplete information. Extensive simulations have confirmed our analytical results.
\end{abstract}


\section{Introduction and motivation}

In 2016, the Japanese Government publicized a bold initiative and a call to action for the implementation of a ``Super Smart Society'' announced as {\em Society 5.0}.
The novelty of  Society 5.0 is that it embodies a sustainable {\em service-centric} society enabled by the latest digital technologies.
Society 5.0 meets the needs of its members by providing goods and services to the people who require them when they are required,
and in the amount required, thus enabling its citizens to live an active and comfortable life through the provisioning of high-quality services 
\cite{deguchi-2020,gladden-2019,hitachi-2020,Shiroishi2018}.
Society 5.0 provides a common societal infrastructure for prosperity based on an advanced service platform which turns out to be its main workhorse.

The insight behind Society 5.0 is that continued progress of ICT and digital technologies of all sorts will provide individuals and society tremendous
opportunities for innovation, growth, and unprecedented prosperity and well-being through various forms of human-to-human, human-to-machine,
and machine-to-machine cooperations and collaboration. Most of these forms of cooperation and collaboration between humans and machines or between autonomous
machine systems have yet to be defined and understood. 

Services and their effects have been studied intensely in the past two decades and most of their dynamics are now well understood 
\cite{Chesbrough2006,Larson2016,Maglio2008,Maglio2009,Medina-Borja2015,Spohrer2007}. Recently, the emergence of Decentralized Autonomous Organizations (DAO) has 
motivated the study of service provisioning in decentralized blockchain-based environments fed by open networks of contributors \cite{bellini-2020,hasan-2022,santana-2022}. 

Our paper was inspired and motivated by some of the challenges that will have to be overcome in order to implement Society 5.0. Key among these challenges, as pointed out by several workers, 
is providing trusted and secure services \cite{kaji-2021,gandini-2016,adebesin-2020}. 
With this in mind, we set out to explore providing a trust and reputation service in recently proposed blockchain-based decentralized marketplaces. To the best of our knowledge,
this is the first time such an effort has been undertaken. 

Today, decentralized markets are growing at a rapid pace with all types of goods and services being transacted online.
In such global markets, buyers and sellers 
engage in transactions with counterparts with whom they had little or no previous interaction. This introduces significant risks for both buyers and sellers. 
In order to assist buyers (sellers) with the process of choosing a trustworthy trading partner, marketplaces maintain 
individual {\em reputation scores} for each seller (buyer) \cite{koutsos-2021,peng-2020,soska-2016,travizano-2018}.
These reputation scores capture, in various forms, statistical information about the past behavior of sellers (buyers) registered with the platform.

The goal of a trust and reputation service is to provide buyers with a robust framework that allows them to select future transaction partners based on a combination
of objective and subjective trust measures distilled from accumulated evidence of sellers’ past behavior in the marketplace. The quality of a trust and reputation service depends, 
in a fundamental way, on the quality of the feedback it receives from buyers. 
This is even more crucial when we consider decentralized marketplaces, where there is no centralized control, unlike marketplaces such as Amazon and eBay.

Being a subjective measure, the quality of buyer feedback is notoriously hard to assess \cite{braga-2018,josang-2002,teacy-2006}.
The fundamental problem is that different buyers may rate a similar experience with the same seller vastly differently. 
When feedback is provided by buyers from around the world, who may value different
aspects of the same transaction differently, it is very hard to know when a buyer provides truthful feedback. 

\subsection{Our contributions}

The first main contribution of this paper is to propose a novel blockchain-based trust and reputation service with the goal of reducing the uncertainty associated with 
buyer feedback in decentralized marketplaces. 

The second main contribution of the paper is to illustrate three applications of the proposed blockchain-based trust and 
reputation service.  Specifically, in Subsections \ref{subsec:price-range} and \ref{subsec:service-type} we discuss two applications of our 
service to a multi-segment marketplace, where a malicious seller may establish an
enviable reputation by selling cheap items or providing some specific service, only to use their superb reputation score to defraud buyers 
in a different market segment. Next, in Subsection \ref{subsec:discount}, we apply the results of
Section \ref{sec:trust-measure} in the context of sellers with time-varying performance due, for example, to overcome initial difficulties. We provide two discounting schemes 
where older reputation scores are given less weight than more recent ones, thus focusing attention on current performance. Finally, in Subsection \ref{subsec:long-term} 
we show how to predict trust and reputation scores far in the future, based on incomplete information.

In our work we assume that a {\em Smart Contract} (SC) is associated with each transaction. 
We assume that the SC in charge of the transaction is also responsible for providing feedback at the end of the
transaction, replacing buyer feedback with a more objective assessment of how well the buyer and the seller have fulfilled their contractual 
obligations towards each other.

At the heart of any trust and reputation service must lie a {\em trust engine}, an algorithm that takes as input a seller's reputation score and distills
from it a subjective trust measure, namely the perceived probability that on the next transaction, the seller will fulfill her contractual obligations.
The proposed trust and reputation engine was inspired by a classic result in probability theory, namely Laplace’s Law of Succession \cite{geisser-1984,zabell-1989}. 
We extend Laplace's classic result in a way that provides a trust measure in a seller's future performance in
terms of her past reputation scores.

The remainder of this paper is organized as follows. 
Section \ref{sec:background} offers a succinct review of recently proposed blockchain-based trust and reputation systems. 
Section \ref{sec:trust-measure} introduces the proposed Laplace trust and reputation service. This is followed by Section \ref{sec:updating-trust} which discusses how the trust measure 
is updated over time.
Section \ref{sec:extensions} offers three applications of the proposed Laplace trust and reputation service. 
Section \ref{sec:simul}  introduces our simulation model and offers simulation results.
Finally, Section \ref{sec:concl} offers concluding remarks and directions for future work. 
\noindent

We wish to alert the reader that an appendix was added for some tedious mathematical derivations whose inclusion in the main paper would be distracting.

\section{Blockchain-based reputation systems}\label{sec:background}

Trust and reputation models have long been of interest to economists 
\cite{bass-1969,bergemann-2018,Frederick2002,Howard1966,lucking-2000,resnick-2006,shapiro-1983,w-1995}.
The advent of e-commerce has renewed interest in online transactions where, naturally, trust or lack thereof is a major concern.

In recent years, a steadily increasing number of workers have investigated blockchain-based reputation systems wherein SCs may or may not play a significant role. We refer the reader
to the  surveys of Hendrix {\em et al.} \cite{hendrix-2014}, Bellini {\em et al.} \cite{bellini-2020}, and Hasan {\em et al.} \cite{hasan-2022} for a comprehensive discussion. 
With this in mind, the main goal of this section is to review some of the recently proposed blockchain-based reputation systems.

Buechler {\em et al.} \cite{buechler-2015} developed a reputation system where SCs contribute to the task of reputation scoring by analyzing the underlying network 
structure. Their system allows buyers and sellers to query and record the outcomes of transactions.

Lu {\em et al.} \cite{lu-2018} proposed a blockchain-based trust model specifically designed to improve the trustworthiness of messages in 
Vehicular Ad-hoc Networks (VANET). 
However, their system does not use SCs in any capacity. Later, Javaid {\em et al.} \cite{javaid-2019} proposed a blockchain-based and a trusted Certificate-Authority-based trust and 
reputation model for VANET. While SCs are mentioned by the authors of \cite{javaid-2019}, no specific role for SCs is mentioned in the paper, other than supporting 
the functionality of the blockchain.  More recently, Singh {\em et al.} \cite{singh-2020} have proposed a blockchain-based trust management system in the context of the
Internet of Vehicles \cite{olariu-2020,olariu-2021}, an extension of VANET. In their work, the blockchain provides trust 
among vehicles that have no reason to trust each other. The blockchain also manages in a reliable manner trust and reputation across the Internet of Vehicles. 
However, although mentioned, there is no specific role played by SCs in their scheme.

Arshad {\em et al.} \cite{arshad-2022} presented a blockchain-based reputation system that they call REPUTABLE which computes the reputation of sellers within a blockchain ecosystem 
through decentralized on-chain and off-chain implementations. REPUTABLE ensures privacy, reliability, integrity, and accuracy of reputation scores, all this with minimal
overhead. In order to facilitate gathering buyer feedback, REPUTABLE employs SCs. However, the SCs are not entrusted with providing feedback on their own.

\section{The assumed blockchain-based decentralized marketplace}\label{sec:our-marketplace}

If a reputation system is to be successful, several conditions must be satisfied: first, the decentralized marketplace must collect, aggregate, and disseminate seller 
reputation scores accurately and in a timely manner; second, buyers provide truthful feedback on their buying experience; and, third, buyers base the choice of 
their future transaction partners (i.e. sellers) solely on reputation scores.

The first and third conditions are relatively easy to enforce or to incentivize. The second condition is far more problematic.
It has been argued that if buyers  consistently provide truthful feedback, isolated interactions
between buyers and sellers take on attributes of long-term relationships and, as a result, the reputation scores tallied by the marketplace
become a high-quality substitute for community-based reputation \cite{resnick-2000}.

It is not surprisingly, therefore, that numerous authors have proposed strategies intended to incentivize truthful 
feedback \cite{jurca-2003,jurca-2004,jurca-2005,zhao-2010}. 

In this work we assume a blockchain-based marketplace similar to \cite{dennis-2015,koutsos-2021,soska-2016,peng-2020,travizano-2018,zhou-2021}, where the transactions 
between buyers and sellers are maintained as individual blocks that, once added to the blockchain, keep immutable information about the transaction. 
We maintain statistical information about the buyers' and sellers' performance as part of the blockchain.

\section{The Laplace trust and reputation service}\label{sec:trust-measure}

The main goal of this section is to introduce our trust and reputation service. 

\subsection{Terminology and definitions}\label{subseq:terminology}

Consider a decentralized marketplace and a new seller $S$ who just joined the marketplace at time $0$. 
We associate with the seller an urn containing an unknown number, $N$, of
balls and an unknown composition, in terms of the number of black
balls it contains. The intention is for the urn of unknown composition to represent the total number of transactions in which seller $S$ will be
involved during her career in the marketplace. Here, each black ball represents a transaction in which seller $S$ has fulfilled her contractual obligations. 

We define the {\em reputation score} of the seller at time $t$ as an ordered triple whose first and second components are, respectively, 
the total number of transactions in which the seller was involved up to time $t$ and the number of transactions in 
which the seller has fulfilled her contractual obligations up to time $t$. The third component is $(0,t)$ or, simply, $t$ if no confusion can arise.

Each transaction in which seller $S$ is involved is associated with a ball extracted from the urn {\em without replacement}. If the extracted ball is black, we say that the seller has
fulfilled her obligations in the corresponding transaction. The motivation for this is that every time a ball is extracted from the urn without replacement, the probability
of obtaining a black ball on the next extraction changes. This is intended to capture, to some extent, the uncertainties and vagaries of seller behavior. 

Let $I$ be the random variable denoting the {\em initial} number of black balls in the urn. Let $H_i=\{I=i\},\ (0 \leq i \leq N)$, be the
{\em hypothesis} that the initial composition of the urn is $(i,N-i)$, 
in other words the urn contains $i$ black balls, while the remaining $N-i$ balls have other colors. 


Since nothing is known {\em \`{a}  priori} about the past history,
skill level, and integrity profile of the seller, it makes sense to assume, as an {\em initial prior}, that all compositions of the urn are 
equiprobable (see \cite{geisser-1984} for a good discussion) and so
\begin{equation}\label{Ci}
\Pr[H_i] = \frac{1}{N+1}.
\end{equation}

We define $\rho_S(0,t)$, the {\em trust  measure} in seller $S$ at time $t$, to be the probability that the seller will fulfill her 
contractual obligations on the next transaction following $t$.
In terms of the underlying urn, this means that the next ball extracted from the urn is black. For example, let $B_0$ be the event that on 
the very {\em first} transaction the seller will fulfill her
contractual obligations. Equivalently, $B_0$ is the event that, on the first extraction a black ball will appear. For reasons that will 
become clear later we write $\rho_S(0,0)$ for $\Pr[B_0]$. It is clear that
\begin{eqnarray}\label{b0}
&& \rho_S(0,0) = \sum_{i=0}^{N} \Pr[B_0 | H_i ] \Pr[H_i] 
         = \frac{1}{N+1} \sum_{i=0}^{N} \frac{i}{N}\ \ \ \mbox{[by (\ref{Ci})]}  \nonumber \\
         &=& \frac{1}{N(N+1)} \sum_{i=0}^{N} i \nonumber \\
         &=& \frac{1}{N(N+1)} \frac{N(N+1)}{2} = \frac{1}{2},
\end{eqnarray}
which makes intuitive sense, since we have no \`{a}  prior knowledge of the seller's past behavior in the marketplace and therefore the trust 
we place in her is $\frac{1}{2}$.

\subsection{Updating the prior}\label{subsec:update-1}

Now, suppose that our seller has accumulated, in the time interval $r([0,t]$, a reputation score of $(n,k,t)$. Recall that this means that out of a total
of $n$ transactions in which the seller was involved up to time $t$, she has fulfilled her obligations in $k$ of them. 
Equivalently, this says that from the urn mentioned above, a sample of $n$ balls was extracted {\em without replacement} and
that $k$ of them were observed to be black. 

In order to update the trust measure in our seller, we need to update our belief in the original composition of the associated urn.
For this purpose, let $A$ be the event that in a sample of $n$ balls extracted without replacement from the urn, $k$ black balls were observed. 
Once the event $A$ is known, we update the prior in a Bayesian fashion by setting
\vspace*{5 mm}
\begin{eqnarray}\label{eq:update-11}
&& \Pr[H_i|n,k] =\Pr[H_i |A] = \frac{\Pr[H_i \cap A]}{\Pr[A]} \nonumber \\
         &=& \frac{\Pr[A | H_i]\Pr[H_i]}{\sum_{j=0}^N \Pr[A | H_j]\Pr[H_j]} 
         = \frac{\Pr[A | H_i]}{\sum_{j=0}^N \Pr[A | H_j]} \ \ \mbox{[by (\ref{Ci}).]} \nonumber \\
         &=& \frac{
                   \frac { {i \choose k} {{N-i} \choose {n-k}}} {{N \choose n}}
                  } 
                  {
                     \sum_{j=0}^N \frac{{{j \choose k}} {{N-j} \choose {n-k}}} {{{N} \choose {n}}}
                  } 
         = \frac{  { {i \choose k} {{N-i} \choose {n-k}}}} {\sum_{j=0}^N  {{j \choose k}} {{N-j} \choose {n-k}}} 
                   \nonumber \\ 
         &=& \frac{
                    \frac { {i \choose k} {{N-i} \choose {n-k}}} {{N \choose n}} 
                  }
                  {
                    \frac{ {{N+1} \choose {n+1}}}  {{{N} \choose {n}} }
                   }
            \ \ \ \mbox{[by (\ref{eq:lovasz-1}) in  the Appendix]} \nonumber \\
           &=& \frac{
                    { {i \choose k} {{N-i} \choose {n-k}}}
                  }
                  {
                     {{N+1} \choose {n+1}}.
                  } 
\end{eqnarray}

To summarize, the expression of the updated prior $\Pr[H_i|n,k]$ reflects our updated belief in the {\em initial} composition of the urn, as a 
result of seeing $k$ black balls out of $n$ balls extracted. In terms of our seller, upon seeing that the seller has fulfilled her obligations 
in $k$ out of the first $n$ transactions, we update the perceived intrinsic performance profile of our seller. 
At the risk of mild confusion, we continue to write $\Pr[H_i]$ for the updated prior, instead of more cumbersome $\Pr[H_i|n,k]$.

\subsection{Modeling the trust measure}\label{subsec:trust-1}

Recall that we define a seller's (subjective) trust measure, $\rho_S(0,t)$, at time $t$ as the probability of the event that on the
next transaction the seller will fulfill her contractual obligations.

\begin{theorem}\label{thm:kn}
Assuming that seller $S$ has accumulated, in the interval $(0,t)$, a reputation score of $(n,k,t)$, the trust measure in $S$ at time $t$ is
\begin{equation*}\label{eq:kn}
\rho_S(0,t) = \frac{k+1}{n+2}.
\end{equation*}
\end{theorem}

\begin{IEEEproof}
Consider the urn associated with seller $S$ and assume that out of the urn, a sample of $n$ balls was extracted and $k$ of them were observed to be black.
Let $B$ be the event that the next ball extracted from the urn is black. In terms of our marketplace, $\Pr[B]$ is precisely $\rho_S(0,t)$.
By the Law of Total Probability,
\begin{equation}\label{eq:B-1}
\Pr[B] = \sum_{i=0}^N \Pr[B |H_i] \Pr[H_i].
\end{equation}
Observe that $\Pr[B | H_i] = \frac {i-k}{N-n}$ and recall that, by (\ref{eq:update-11}), 
$\Pr[H_i] = \frac { {i \choose k} {{N-i} \choose {n-k}}} {{{N+1} \choose {n+1}}}.$
With this, (\ref{eq:B-1}) can be written as
\begin{eqnarray*}\label{eq:B-11}
\rho_S(0,t)= \Pr[B] 
       &=& \sum_{i=0}^N \frac{i-k}{N-n} \frac { {i \choose k} {{N-i} \choose {n-k}}} {{{N+1} \choose {n+1}}}  \nonumber \\
       &=& \frac{\sum_{i=0}^N (i-k) \cdot \frac{i!}{k! (i-k)!} {{N-i} \choose {n-k}}} {(N-n) \frac{(N+1)!}{(n+1)! (N-n)!}}  \nonumber \\
       &=& \sum_{i=0}^N \frac{ (k+1) {i \choose {k+1} } {{N-i} \choose {n-k}} } { (N+1) {{N} \choose {n+1}}} \nonumber \\
       &=& \frac{k+1}{N+1} \sum_{i=0}^N \frac { {i \choose {k+1}} {{N-i} \choose {n-k}}} {{{N} \choose {n+1}}} \nonumber \\
       &=& \frac{k+1}{N+1}~ \frac{ {{N+1} \choose {n+2}}}{{{N \choose {n+1}}}} \ \ \ \mbox{[by (\ref{eq:lovasz-1})]} \nonumber \\
       &=& \frac{k+1}{n+2},
\end{eqnarray*}
and the proof of Theorem \ref{thm:kn} is complete.
\end{IEEEproof}

\subsection{Illustrating Theorem \ref{thm:kn}}\label{subsec:them}
Somewhat surprisingly, the expression of the trust measure is independent of $N$ and depends only on $n$ and $k$. 
It is very important to note that the expression of the trust measure specified in Theorem \ref{thm:kn} is very easy to remember and to compute. 
Specifically, if a certain seller has accumulated a reputation score $(n,k,t)$, evaluating the corresponding trust measure in the seller at time $t$
is very simple. This is one of the significant advantages of our trust and reputation service.

\begin{figure}[!h]
\centering
\includegraphics[clip, scale=1.5]{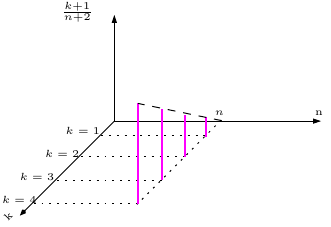}
\caption{\em Illustrating the trust measure for fixed $n$.}
\label{fig:fixed-n}
\end{figure}

It is of interest to plot the trust measure of Theorem \ref{thm:kn} for the cases where $n$, and $k$ are fixed. First, for fixed $n$,
Figure \ref{fig:fixed-n} reveals that $\rho_S(0,t)$ features a {\em linear} increase in $k$. Indeed, the trust measure of a seller with a reputation
score of $(n,k+1,t)$ and that of a seller with a reputation score of $(n,k,t)$, differ by $\frac{k+2}{n+2} - \frac{k+1}{n+2} = \frac{1}{n+2}$.

\begin{figure}[!h]
\centering
\includegraphics[clip, scale=1.5]{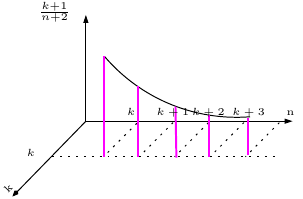}
\caption{\em Illustrating the trust measure for fixed $k$.}
\label{fig:fixed-k}
\end{figure}

On the other hand, Figure \ref{fig:fixed-k} shows that for a {\em fixed} value of $k$, $\rho_S(0,t)$, perceived as a function of $n$,
experiences a {\em hyperbolic} decline. To see this, observe that the trust measure of a seller with a reputation
score of $(n,k,t)$ and that of a seller with a reputation score of $(n+1,k,t)$, differ by 
$\frac{k+1}{n+2} - \frac{k+1}{n+3} = (k+1) \left [ \frac{1}{n+2} - \frac{1}{n+3} \right ]$.

\begin{figure}[!h]
\centering
\includegraphics[clip, scale=0.16]{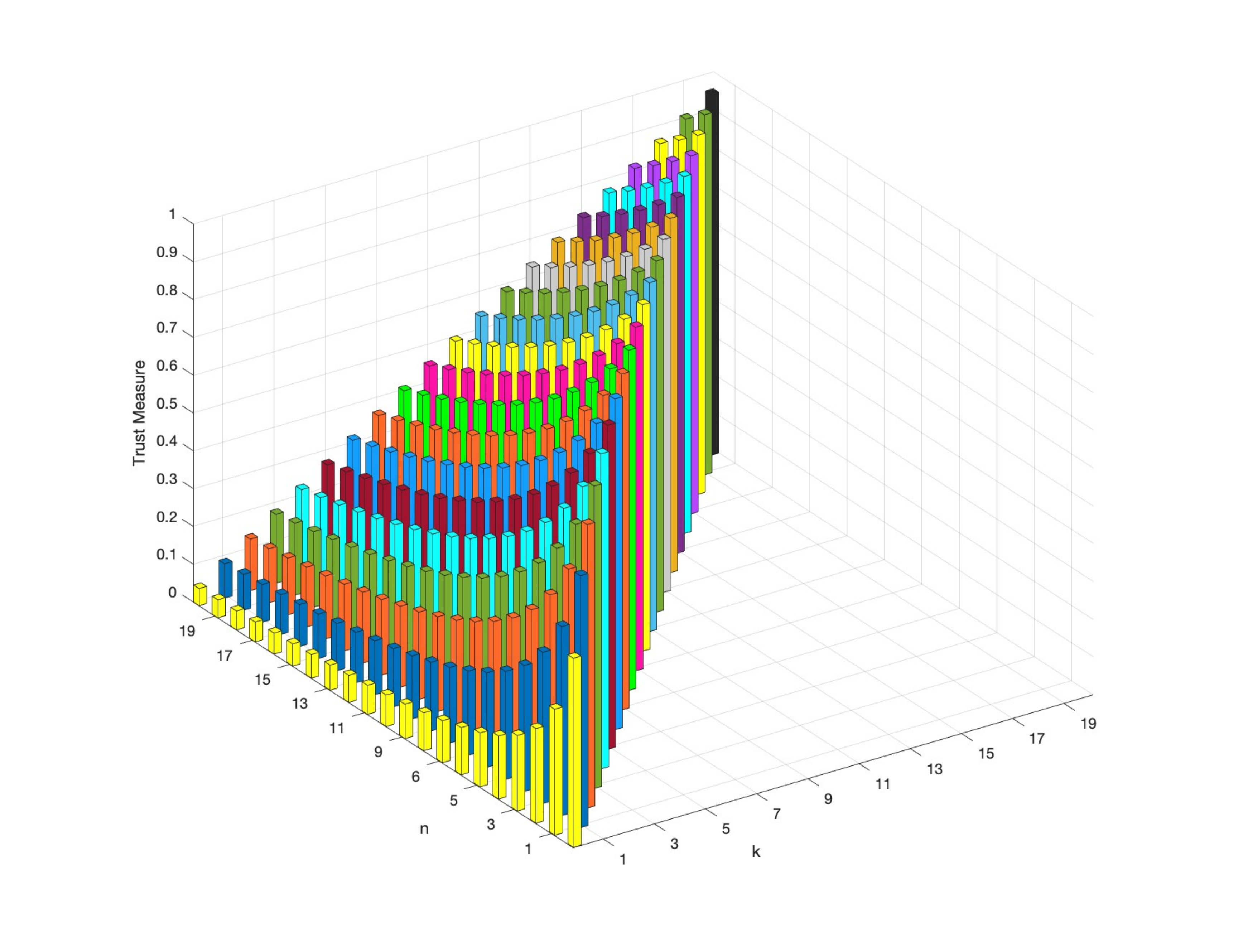}
\caption{\em Illustrating $\rho_S(0,t)$ for small values of $n$ and $k$}
\label{fig:kn}
\end{figure}

\vspace*{5 mm}
To summarize this discussion, we refer the reader to Figure \ref{fig:kn} the trust measure $\rho_S(0,t)$ for small values of $n$ and $k$.
For a better illustration, the values of $\rho_S(0,t)$ for different values of $k$ are depicted in different colors. 
Figure \ref{fig:kn} also reveals that $\rho_S(0,0)=\frac{1}{2}$, as we found in (\ref{b0}).

\section{Updating the trust measure}\label{sec:updating-trust}

The main goal of this section is to show how the trust measure introduced in Section \ref{sec:trust-measure} is updated over time.

\begin{theorem}\label{thm:k'n'}
Assume that in the time interval $(0,t]$, seller $S$ was involved in $n$ transactions and that she has fulfilled her contractual obligations in $k$ of them. If in the time interval 
$(t,t']$ seller $S$ is involved in $n'$ additional transactions and that she fulfills her
contractual obligations in $k'$ of them, then the seller's trust measure, $\rho_S(0,t')$, at time $t'$ is
\begin{equation}\label{eq:k'n'}
\rho_S(0,t') = \frac{k+k'+1}{n+n'+2}.
\end{equation}
\end{theorem}

\begin{IEEEproof}
Let $A'$ be the event that in a subsequent sample of size $n'$, $k'$ balls were observed to be black. Once the event $A'$ is known to have occurred,
it is necessary to update our prior. Proceeding, in a Bayesian fashion, we write
\begin{eqnarray}\label{eq:update-2}
\Pr[H_i|n,k,n',k'] &\leftarrow& \Pr[H_i |A'] = \frac{\Pr[H_i \cap A']}{\Pr[A']} \nonumber \\
         &=& \frac{\Pr[A' | H_i]\Pr[H_i]}{\sum_{j=0}^N \Pr[A' | H_j]\Pr[H_j]}.
\end{eqnarray}


Noticing that
\begin{itemize}
\item $\Pr[A' | H_i] = \frac { {i \choose k} {{N-i} \choose {n-k}}} {{N \choose n}}$;
\item  by (\ref{eq:update-11}), $\Pr[H_i] = \frac{ { {i \choose k} {{N-i} \choose {n-k}}} } { {{N+1} \choose {n+1}}}$; and,
\item by (\ref{A'-1}) in the Appendix  \ref{eval-A'}, $\Pr[A']=\sum_{j=0}^N \Pr[A' | H_j]\Pr[H_j] =
\frac{ {{k+k'} \choose k} {{n-k + n'-k'} \choose {n-k}} } { {{n+n'+1} \choose {n+1}}}$,
\end{itemize}
equation (\ref{eq:update-2}) becomes
\begin{equation}\label{eq:update-21}
\Pr[H_i] = \Pr[H_i|n,k,n',k'] = \frac { {i \choose {k+k'}} {{N-i} \choose {n-k +n'-k'}}} {{{N+1} \choose {n+n'+1}}}.
\end{equation}

As before, in order to simplify notation, we continue to refer to $\Pr[H_i|n,k,n',k']$ as $\Pr[H_i]$.
The expression of the prior $\Pr[H_i]$ in (\ref{eq:update-21}) reflects our updated belief in the composition of the urn, as a result of seeing $k'$ black balls
out of $n'$ balls in the second sample extracted. 


Let $B'$ be the event that the next ball extracted from the urn is black. In terms of our marketplace, $\Pr[B']$ is $\rho_S(0,t')$. 
\begin{eqnarray}\label{eq:B'-11}
 \Pr[B'] &=& \sum_{i=0}^N \Pr[B' |H_i] \Pr[H_i] \nonumber \\
       &=& \sum_{i=0}^N \frac{i-k-k'}{N-n-n'} \frac { {i \choose {k+k'}} {{N-i} \choose {n-k+n'-k'}}} {{{N+1} \choose {n+n'+1}}}  \nonumber \\
       &=& \scriptstyle  \frac{1} {(N-n-n') {{N+1} \choose {n+n'+1}}}   \sum_{i=0}^N (i-k-k') \cdot \frac{i!}{(k+k')! (i-k-k')!} {{N-i} \choose {n-k+n'-k'}} \nonumber \\
       &=& \scriptstyle  \frac{k+k'+1} {(N-n-n') {{N+1} \choose {n+n'+1}}}   \sum_{i=0}^N {i \choose {k+k'+1} } {{N-i} \choose {n-k+n'-k'}} \nonumber \\
       &=& \frac{k+k'+1} {(N-n-n') {{N+1} \choose {n+n'+1}}} {{N+1} \choose {n+n'+2}} \nonumber \\
       &=& \frac{k+k'+1}{n+n'+2}.
\end{eqnarray}
\end{IEEEproof}

Notice that, in spite of the laborious derivation, the final result is extremely simple and {\em easy} to compute. This is a definite advantage of our scheme.

An interesting question is to determine under what conditions the trust measure $\rho_S(0,t')$ is at least as large as $\rho_S(0,t)$. The answer to this question is provided 
by the following result.
\begin{lemma}\label{better}
$$
\rho_S(0,t') \geq \rho_S(0,t) \iff \frac{k'}{n'} \geq \frac{k+1}{n+2}.
$$
\end{lemma}
\begin{IEEEproof}
Follows by Lemma \ref{lem:aa'bb'} in the appendix $a=k+1$, $b=n+2$, $a'=k'$ and $b'=n'$.
\end{IEEEproof}

\begin{figure}[!h]
\centering
\includegraphics[clip, scale=1.5]{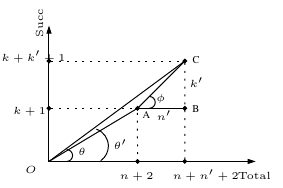}
\caption{\em A geometric interpretation of Lemma  \ref{better}.}
\label{fig:better}
\end{figure}
\vspace*{5 mm}
Refer to Figure \ref{fig:better} for a geometric illustration of Lemma \ref{better}.
Consider a two-dimensional coordinate system where the horizontal and vertical axes capture, respectively, the total number of transactions and the number of transactions in which the seller 
has fulfilled her contractual obligations. Consider, further, the points $A,\ B,\ C$ of coordinates $(n+2,k+1),\ (n+n'+2,k+1),\ (n+n'+2,k+k'+1)$. It is easy to confirm that
$\rho_S(0,t) = \tan \theta = \frac{k+1}{n+2}$, and $\rho_S(0,t') = \tan \theta' = \frac{k+k'+1}{n+n'+2}$. Finally, is is easy to confirm that $\rho_S(0,t') \geq \rho_S(0,t)$ 
if and only if the angle $\phi$ determined by the sides $AB$ and $AB$ of the triangle determined by the points $A,\ B,\ C$
satisfies $\frac{k'}{n'} = \tan \phi \geq \tan \theta = \frac{k+1}{n+2}$, exactly as claimed in Lemma \ref{better}. 
\vspace*{2mm}

\noindent
Theorem \ref{thm:k'n'} can be readily generalized.
\begin{theorem}\label{thm:gen}
For an arbitrary positive integer $r$, consider $r$ successive epochs $(t_0,t_1],\ (t_1,t_2],\ \cdots,\ (t_{i-1},t_i],\ \cdots,\ (t_{r-1},t_r]$,
such that in epoch $(t_{i-1},t_i],\ (1 \leq i \leq r)$, our seller was involved in $n_i$ transactions and has fulfilled her contractual obligations in $k_i$ of them.
Then the seller's reputation score at time $t_r$ is $(\sum_{i=1}^r n_i, \sum_{i=1}^r k_i, (t_0,t_r))$
and her associated trust measure is
\begin{equation}\label{eq:tr-trust}
\rho_S(0,t_r) = \frac{\sum_{i=1}^r k_i +1}{\sum_{i=1}^r n_i+2}.
\end{equation}
\end{theorem} 
\begin{IEEEproof}
Assume, without loss of generality, that $t_0=0$ and let $t$ and $t'$ denote, respectively, $t_{r-1}$ and $t_r$. In the time interval $(0,t]$ the seller was involved in 
$\sum_{i=1}^{r-1} n_i$ transactions and has fulfilled her obligations in $\sum_{i=1}^{r-1} k_i$ of them. In the time interval, $(t,t']$ our seller was involved in
$n_r$ transactions and has fulfilled her contractual obligations in $k_r$ of them.

By definition, in the interval $(0,t']$, the seller's reputation score is $(\sum_{i=1}^r n_i, \sum_{i=1}^r k_i, (t_0,t_r))$. Similarly, by Theorem \ref{thm:k'n'}
her trust measure is
\begin{eqnarray*}
\rho_S(0,t_r) &=& \frac{k+k'+1}{n+n'+2} \nonumber \\
              &=& \frac{ (\sum_{i=1}^{r-1} k_i) + k_r + 1}{(\sum_{i=1}^{r-1}  n_i) + n_r +2} \nonumber \\
              &=& \frac{\sum_{i=1}^r k_i +1}{\sum_{i=1}^r n_i+2},
\end{eqnarray*}
and the proof of Theorem \ref{thm:gen} is complete.
\end{IEEEproof}

Theorem \ref{thm:gen} has a number of consequences:
\begin{itemize}
\item the updated trust measure is related to the updated reputation scores, exactly as specified in Theorem \ref{thm:kn};
\item the updated trust measure does not change if
   \begin{itemize}
      \item {\bf Associativity:} the seller has fulfilled her obligations in $0$ of the first $\sum_{i=1}^{r-1} n_i$ transactions and in $\sum_{i=1}^r k_i$ out of the next $n_r$
       transactions, provided $\sum_{i=1}^r k_i \leq n_r$.
      \item {\bf Commutativity:} for any choice of subscripts $i, j$, with $(1 \leq i \neq j \leq r)$, the $n_j$ transactions in epoch $j$ have occurred before or 
after the $n_i$ transactions in epoch $i$;
      \item {\bf Interchangeability:} the seller has fulfilled her obligation in $k_j$ of the $n_i$ transactions in epoch $i$ and in $k_i$ of the transactions in epoch $j$, 
      provided that $k_j \leq n_i$ and $k_i \leq n_j$. 
    \end{itemize}
\end{itemize}

\section{Applications of the Laplace Trust Engine}\label{sec:extensions}

The main goal of this section is to illustrate three applications of the trust and reputation service introduced in Section \ref{sec:trust-measure}. 
Specifically, in Subsections \ref{subsec:price-range} and \ref{subsec:service-type} we discuss two applications to a multi-segment marketplace, 
where a malicious seller may establish a very high reputation by selling cheap items or providing some specific type of service, only to use 
their reputation score to defraud buyers in a different market segment. 

Next, in Subsection \ref{subsec:discount}, we apply the results of Section \ref{sec:trust-measure} in the context of sellers with time-varying 
performance due to an initial learning curve. We provide two discounting schemes, wherein older reputation scores are
given less weight than more recent ones. Finally, in Subsection \ref{subsec:long-term} we show how to predict trust and reputation scores 
far in the future, based on currently available information.

\subsection{Price-range specific trust and reputation}\label{subsec:price-range}

We assume that the transactions in the marketplace are partitioned, by the monetary value of the goods transacted, into non-overlapping price ranges 
$0 < R_1 < R_2 < \cdots < R_s$ for some positive integer $s$. These ranges determine $s$ {\em market segments} $M_1, M_2, \cdots, M_s$ where 
market segment $M_j$ involves all the transactions within the price range $R_j$.

In all marketplaces of which we are aware 
\cite{bellini-2020,hasan-2022,gandini-2016,adebesin-2020,koutsos-2021,peng-2020,soska-2016,travizano-2018,josang-2002,teacy-2006}, 
seller reputation is {\em global}, being established irrespective of their performance in different market segments.

However, this may lead to insecurities. For example, imagine a seller that has 
established an enviable reputation score by selling cheap items, all in the market segment corresponding to the range $R_1$. Suppose that
our seller decides to get involved in a different market segment, say corresponding to price range $R_{10}$. Should her reputation score 
established in $R_1$ carry over to $R_{10}$? We believe that the answer should be in the negative. One reason is that, as pointed out 
by \cite{kerr-2009} and other workers, dishonest sellers establish stellar reputation scores by selling cheap items and use the resulting reputation score 
to {\em hit-and-run} in a different market segment. 

To prevent this kind of attack from being mounted, we associate with each market segment a distinct reputation score and, consequently, a distinct 
trust measure. Also, with each market segment, we associate a different urn as discussed in the previous sections of this work.
For example, if our seller has never transacted in the market segment corresponding to the price range $R_{10}$, her reputation score in 
that market segment is $(0,0,t)$ and, not surprisingly, her corresponding trust measure will be $\frac{0+1}{0+2} = \frac{1}{2} = 50\%$, capturing 
the idea that nothing is known about the performance of the seller in that market segment.

Consider a generic market segment  $M_i,\ (1 \leq i \leq s)$, and assume that up to time $t$, our seller has accumulated a reputation score of 
$(n_i,k_i,t)$ in $R_i$. Consistent with our definition, the trust measure that our seller enjoys in $M_i$ is $\frac{k_i+1}{n_i+2}$. This trust 
measure is {\em local} to $M_i$ and is independent of the seller's trust measure in other market segments. 

It is worth noting that, as an additional benefit, our approach provides {\em resistance} to Sybil attacks. It is well known that malicious users 
involve their Sybils in augmenting their reputation scores \cite{cheng-2005,nasrulin-2022,santana-2022,stannat-2021}. However, the fact that 
by assumption Smart Contracts are responsible for providing transaction feedback (including the market segment in which the transaction 
took place), this feedback will be, perforce, local to one market segment, minimizing the effect of the attack. Indeed, as a result of the 
Sybil attack, the malicious user's reputation may well increase in one market segment, 
but her reputation in other market segments will not be affected. This provides for very desirable resistance to Sybil attacks.

\subsection{Service-specific trust and reputation}\label{subsec:service-type}

In Subsection \ref{subsec:price-range} we argued that reputation scores and, therefore, the trust measure of a seller should not be global but should, 
instead, be specific to individual price ranges. Specifically, we made the point that reputation scores acquired by doing business in one market 
segment (by dollar amount) should not carry over to a different market segment.

In this subsection, we extend the same idea to the types of services provided. The intuition is that a service provider (i.e. seller) may behave 
differently when providing different services. Thus, the best indicator of how the service provider will perform in the future depends on their 
past performance in the context of the type of services contemplated. This motivates assessing the trustworthiness of a service provider by 
the type of individual service of interest.

As an illustrative example, consider a plumbing contractor who may act in the marketplace as a seller of plumbing hardware, but also as a provider of 
plumbing services such as repairs, installation of various equipment such a gas furnaces, electric furnaces, hot water heaters, or extended 
maintenance contracts, etc.

Our plumber may be inclined to provide higher quality services in areas that benefit him most (e.g. installing electric water heaters) and of 
lesser quality in some other areas that are less lucrative, e.g. maintenance contracts or installing gas water heaters), even though an electric 
water heater may cost roughly the same as a gas water heater. 

The point is that the plumber's  reputation score acquired by providing one type of service should not be relevant when evaluating his/her 
trustworthiness in different service categories where he/she is either less competent or simply not interested in providing high-quality services.

\subsection{Discounting old trust measures}\label{subsec:discount}

Up to this point, we have assumed that seller behavior is constant over time. For various reasons, sellers may well change their attitude and behave differently 
from the way they acted in the past. To accommodate this imponderable, in this subsection we introduce two simple mechanisms that allow us to discount older trust measures,
giving more credence to recent reputation scores.

For an arbitrary integer $r$, consider $r$ successive time epochs $(t_0,t_1],\ (t_1,t_2],\ \cdots,\ (t_{i-1},t_i],\ \cdots,\ (t_{r-1},t_r]$
with $t_0=0$ and such that in epoch $(t_{i-1},t_i],\ (1 \leq i \leq r)$, our seller was involved in $n_i$ transactions and has fulfilled her contractual obligations in $k_i$ of them.
Recall that, given this information, the seller's reputation score at time $t_r$ is 
$(\sum_{i=1}^r n_i,  \sum_{i=1}^r k_i, (t_0,t_r))$ and, by Theorem \ref{thm:gen},  her associated trust measure is
\begin{equation}\label{eq:tr-trust-1}
\rho_S(0,t_r) = \frac{\sum_{i=1}^r k_i +1}{\sum_{i=1}^r n_i+2}.
\end{equation}

\noindent
\subsubsection{First discounting scheme}\label{first}

In order to produce a weighted version of (\ref{eq:tr-trust-1}), consider non-negative rational numbers $\lambda_1,\ \lambda_2,\ \cdots, \lambda_r$
Assuming $\sum_{1 \leq j \leq r} \lambda_j \neq 0$, define weights $0 \leq w_1,\ w_2,\ \cdots,\ w_r \leq 1$,  where, for all $i,\ (1 \leq i \leq r)$,
\begin{equation*}
w_i = \frac{\lambda_i}{\sum_{j=1}^r \lambda_j}.
\end{equation*}


\noindent
Assuming $\sum_{1 \leq j \leq r} \lambda_i \neq 0$, define the following weighted version of (\ref{eq:tr-trust-1}):
\begin{eqnarray}\label{first-weighted}
 D_S(0,t_r) &=& \frac{\sum_{i=1}^r w_i k_i +1}{\sum_{i=1}^r w_i n_i+2} \nonumber \\
            &=& \frac{\sum_{i=1}^r \frac{\lambda_i k_i}{\sum_{j=1}^r \lambda_j} +1}{\sum_{i=1}^r \frac{\lambda_i n_i}{\sum_{j=1}^r \lambda_j}+2} \nonumber \\ 
            &=& \frac{\sum_{i=1}^r \lambda_i k_i + \sum_{i=1}^r \lambda_i}{ \sum_{i=1}^r \lambda_i k_i + 2 \sum_{i=1}^r \lambda_i} \nonumber \\
            &=& \frac{ \sum_{i=1}^r \lambda_i (k_i+1) }{ \sum_{i=1}^r \lambda_i (n_i+2)}.
\end{eqnarray} 

It is clear that by varying the $\lambda_i$s we can give different weights to the past versus more recent trust measures of the seller.
For example, by taking $\lambda_i =0$ for $1 \leq i \leq r-1$ and $\lambda_r=1$, the past performance of the seller is ignored and
the discounted trust measure $D_S(0,t_r)$ reduces to the most recent trust measure. 
Conversely, by taking $\lambda_i =1$, for $(1 \leq i \leq r-1)$, and $\lambda_r=0$, the past is given more weight to the detriment of the more recent performance.
We claim that
\begin{lemma}\label{lem:d1}
\begin{equation}\label{discount-ineq-D}
\min_{1 \leq i \leq r} \frac{k_i+1}{n_i+2}  \leq D_S(0,t_r) \leq \max_{1 \leq i \leq r} \frac{k_i+1}{n_i+2},
\end{equation}
\end{lemma}
\begin{IEEEproof}
Assume, without loss of generality, that $\frac{a_j+1}{b_j+2} = \max_{1 \leq i \leq r} \frac{k_i+1}{n_i+2}$. The proof of the rightmost inequality in the chain above follows
directly from Lemma \ref{lem:a0b0} in the appendix file by setting $a_0=k_j+1$, $b_0=n_j+2$, and 
for all $i,\ (1 \leq i \leq r)$, $a_i= k_i+1$ and  $b_i=n_i+2$.
The leftmost inequality is followed by a mirror argument.
\end{IEEEproof}
 
Lemma \ref{lem:d1} shows that the discounted trust measure {\em cannot} improve the overall trust measure. It can, however, focus attention to more recent
performance that, in many contexts, may be more relevant.

\noindent
\subsubsection{Second discounting scheme}\label{second}

We find it useful to inherit the notation and terminology developed in the previous subsection. In order to produce a simple discounting scheme, we compute a weighted average
of the seller trust measure in each of the $r$ time epochs.
Indeed, consider the time interval $(0,t_r]$ during which a seller has been active in the marketplace. Let $0=t_0 < t_1 < t_2 < \cdots < t_{r-1} < t_r $
be an arbitrary partition of $(0,t_r]$. Assume, further, that for all $i,\ (1 \leq i \leq t)$, in the time interval $(t_{i-1},t_i]$ the seller has 
accumulated a reputation score $(n_i,n_i,t_i-t_{i-1})$. Define non-negative weights $w_1,\ w_2,\ \cdots,\ w_r$,  with $\sum_{i=1}^r w_i =1$, and
define the {\em weighted} trust measure of the seller in the time interval $(t_0,t_r]$ as
\begin{equation}\label{D'}
D'_S(0,t_r) = \sum_{i=1}^r w_i \times \frac{k_i+1}{n_i+2}.
\end{equation}

It is clear that by varying the $w_i$s we can give different weights to the past versus more recent trust measure of the seller. 
It is also easy to see that the following result holds, mirroring (\ref{discount-ineq-D}).
\begin{lemma}\label{lem:d2}
\begin{equation}\label{discount-ineq-D'}
\min_{1 \leq i \leq r} \frac{k_i+1}{n_i+2}  \leq D'_S(0,t_r) \leq \max_{1 \leq i \leq r} \frac{k_i+1}{n_i+2},
\end{equation}
\end{lemma}
\begin{IEEEproof}
We prove that $D'_S(0,t_r) \leq \max_{1 \leq i \leq r} \frac{k_i+1}{n_i+2}$. The proof that $\min_{1 \leq i \leq r} \frac{k_i+1}{n_i+2}  \leq D'_S(0,t_r)$ is similar and, therefore, omitted.
As before, assume without loss of generality that $\frac{a_j+1}{b_j+2} = \max_{1 \leq i \leq r} \frac{k_i+1}{n_i+2}$. With this assumption, we write
\begin{eqnarray}\label{second-weighted}
D'_S(0,t_r) &=& \sum_{i=1}^r w_i \times \frac{k_i+1}{n_i+2} \nonumber \\
            &\leq& \sum_{i=1}^r w_i \times \frac{k_j+1}{n_j+2} \nonumber \\
            &=& \frac{k_j+1}{n_j+2} \sum_{i=1}^r w_i \nonumber \\
            &=& \frac{k_j+1}{n_j+2}.
\end{eqnarray}
This completes the proof of Lemma \ref{lem:d2}.
\end{IEEEproof}

Just like Lemma \ref{lem:d1}, Lemma \ref{lem:d2} tells us that the discounted trust measure {\em will not} improve the overall trust measure. It can, and does, focus attention on more recent
performance that, in many contexts, may be of more relevance.

\subsection{Predicting trust measure and reputation scores over the long term}\label{subsec:long-term}

It is of great theoretical interest and practical relevance to be able to extrapolate the correct performance of a seller and predict her performance, far in the future. With this in mind,
consider a seller that has completed $n$ transactions and has fulfilled her obligations in $k$ of them. Let $A$ be the corresponding event.
We are interested in predicting the {\em expected} reputation score of 
the seller by the time her total number of transactions has reached $n+m$ for some $m \geq 0$.

Let $R$ be the random variable that keeps track of the number of black balls among the additional $m$ balls extracted, and assume that the event $\{R=r\}$ has
occurred.

Using the expression of $H_i$ from (\ref{eq:update-11}), the conditional probability of the event $\{R=r\}$ given $A$ is
\begin{eqnarray}\label{K=k'}
\Pr[R=r|A] &=&  \sum_{i=0}^N \Pr[R=r| H_i] \Pr[H_i] \nonumber \\
	  &=&\frac{ {{k+r} \choose k} {{n-k + m-r} \choose {n-k}} } {{{n+m+1} \choose {n+1}}}. 
\end{eqnarray}
Actually, this follows directly from (\ref{A'-2}) in Subsection \ref{eval-A'} of the Appendix by taking $r=k'$ and $m=n'$.

We are interested in evaluating the {\em conditional expectation}, $E[R|A]$, of $R$ given $A$.  
For this purpose, using the Law of Total Expectation, we write
\begin{eqnarray*}
E[R|A] \hspace*{-2mm} &=& \hspace*{-2mm} \sum_{r=0}^{m} r \Pr[R=r|A] \nonumber \\
    \hspace*{-2mm}  &=& \hspace*{-2mm} \sum_{r=0}^{m} r \cdot \frac{ {{k+r} \choose k} {{n-k + m-r} \choose {n-k}}} { {{n+m+1} \choose {n+1}}} 
       \ \ \ \mbox{[By (\ref{K=k'})]} \nonumber \\
 \hspace*{-2mm}  &=& \hspace*{-2mm} \sum_{r=0}^{m} \left [ (k+r+1) - (k+1) \right ] \cdot \frac{ {{k+r} \choose k} {{n-k + m-r} \choose {n-k}}} { {{n+m+1} \choose {n+1}}} \nonumber \\
  \hspace*{-2mm} &=& \hspace*{-2mm} \sum_{r=0}^{m} (k+r+1) \frac{ {{k+r} \choose k} {{n-k + m-r} \choose {n-k}}} { {{n+m+1} \choose {n+1}}} \nonumber \\ 
      &-& \sum_{r=0}^{m} (k+1) \frac{ {{k+r} \choose k} {{n-k + m-r} \choose {n-k}}} { {{n+m+1} \choose {n+1}}} \nonumber \\
 \hspace*{-2mm}  &=& \hspace*{-2mm} \sum_{r=0}^{m} (k+r+1) \frac{ {{k+r} \choose k} {{n-k + m-r} \choose {n-k}}} { {{n+m+1} \choose {n+1}}} \nonumber \\
      &-& (k+1) \sum_{r=0}^{m} \frac{ {{k+r} \choose k} {{n-k + m-r} \choose {n-k}}} { {{n+m+1} \choose {n+1}}}.
\end{eqnarray*}

The two sums will be evaluated separately. We begin by evaluating the following sum:
\begin{eqnarray*}\label{second-sum}
\sum_{r=0}^{m} \frac{ {{k+r} \choose k} {{n-k + m-r} \choose {n-k}}} { {{n+m+1} \choose {n+1}}} &=&
\sum_{r=0}^{m} \frac{ {{k+r} \choose k} {{n-k + m-r} \choose {n-k}}} { {{n+m+1} \choose {n+1}}} \nonumber \\
&=& \frac{ \sum_{r=0}^{m} {{k+r} \choose k} {{n-k + m-r} \choose {n-k}}} { {{n+m+1} \choose {n+1}}} \nonumber \\
&=& \frac{ {{n+m+1} \choose {n+1}}}  {{ {{n+m+1} \choose {n+1}}}} = 1. 
\end{eqnarray*}
This implies that the second sum is $k+1$.
Next, to evaluate the first sum, we notice that
\begin{eqnarray}\label{first-sum}
(k+k'+1) {{k+k'} \choose k} &=& \frac{k+1}{u+1} (k+k'+1) \frac{(k+k')!}{k! k'!} \nonumber \\
   &=& (k+1) \frac { (k+k'+1)!} {(k+1)! k'!} \nonumber \\
   &=&  {{ k+k'+1} \choose {k+1}}.
\end{eqnarray}

Using (\ref{first-sum}), the first sum can be written as
\begin{eqnarray}\label{first-sum-1}
&& \sum_{r=0}^{m} (k+r+1) \frac{ {{k+r} \choose k} {{n-k + m-r} \choose {n-k}}} { {{n+m+1} \choose {n+1}}} \nonumber \\
&=& \frac{k+1} { { {{n+m+1} \choose {n+1}}}}  \sum_{r=0}^{m} {{k+r+1} \choose {k+1}} { {n-k + m-r} \choose {n-k} } \nonumber \\
&=& \frac{k+1} { { {{n+m+1} \choose {n+1}}}} { {n+m+2} \choose {n+2} } \nonumber \\
&=& (k+1) \frac{n+m+2}{n+2}.
\end{eqnarray}
 
By combining the intermediate results developed above, the expression of $E[R|A]$ becomes
\begin{eqnarray}\label{E-final}
E[R|A] &=& \sum_{r=0}^{m} (k+r+1) \frac{ {{k+r} \choose k} {{n-k + m-r} \choose {n-k}}} { {{n+m+1} \choose {n+1}}} \nonumber \\
     &-& (k+1) \sum_{r=0}^{m} \frac{ {{k+r} \choose k} {{n-k + m-r} \choose {n-k}}} { {{n+m+1} \choose {n+1}}} \nonumber \\
  &=& (k+1) \frac{n+m+2}{n+2} - (k+1) \nonumber \\
  &=& (k+1) \left [ \frac{n+m+2}{n+2} -1 \right ] \nonumber \\
  &=& m \cdot \frac{k+1}{n+2}.
\end{eqnarray}  

\noindent

The intuition behind this simple result is as follows: since nothing is known 
about the future, in each of the $m$ hypothetical extractions from the urn, the {\em success} probability, is the same, namely, $\frac{k+1}{n+2}$. Thus, by a well-known result,
the expectation of the number of successes must be $m \cdot \frac{k+1}{n+2}$. 

Let us translate (\ref{E-final}) into the language of trust and reputation. Consider a seller with a current reputation score of $(n,k,t)$. We are interested in 
predicting the reputation score of the seller by time $T$ when her total number of transactions has reached $n+m$. 
By (\ref{E-final}), it follows that out of a total of $n+m$ transactions, the {\em predicted} number of transactions in which our seller has 
fulfilled her obligations is $k + m \frac{k+1}{n+2}$. 

To  put it differently, the {\em expected reputation score} of the seller by time $T$, when she 
was involved in $n+m$ transactions, 
is $(n+m, k+ m\frac{k+1}{n+2}, T)$. Interestingly, as the following derivation shows, the seller's predicted trust measure at time $T$ is still $\frac{k+1}{n+2}$. 
\begin{eqnarray}\label{future}
\rho_S(0,T) &=& \frac{ k + m \frac{k+1}{n+2} +1}{n+m+2} \nonumber \\
          &=& \frac{k(n+2) + m(k+1) + n+2}{(n+2)(n+m+2)} \nonumber \\
          &=& \frac{k(n+m+2) + n+m+2}{(n+2)(n+m+2)} \nonumber \\
          &=& \frac{(k+1)(n+m+2)}{(n+2)(n+m+2)} \nonumber \\
          &=& \frac{k+1}{n+2}.
\end{eqnarray}

\section{Simulation results}\label{sec:simul}

The goal of this section is to present the results of our empirical evaluation of the trust and reputation service discussed, analytically, 
in Sections \ref{sec:trust-measure} -- \ref{sec:extensions}. 

\subsection{Simulation model}\label{subsec:model}

For the purpose of empirical evaluation, we have simulated a blockchain-based decentralized marketplace with SC support. The actors in the 
marketplace are the buyers and the sellers. We assume that a SC is associated with each transaction and, for simplicity, that each transaction
involves one buyer and one seller. The SC in charge of the transaction is responsible for providing feedback at the end of the transaction, 
replacing notoriously unreliable buyer feedback with a more objective assessment of how well the buyer and the seller have fulfilled 
their contractual obligations towards each other. 

The marketplace simulation model consists of a seller who was involved in transactions with multiple buyers. Each transaction can be either 
successful (indicating that the seller has fulfilled her contractual obligations) or failed otherwise. In the simulation, we tracked the number 
of successful transactions and the total number of transactions. The probability 
of a successful transaction is determined based on the goals of the experiment as we explain in the following subsections. For each 
goal, we repeated the experiment a large number of times, as needed.

The remainder of this section is structured as follows. 
In Subsection \ref{subsec:price-multi-segment} we turn our attention to a multi-segment marketplace (by dollar value of the goods transacted) 
and illustrate, by simulation, the reputation scores and trust measure of a generic seller in these market segments.
Next, in Subsection \ref{subsec:service-multi-segment} we present simulation results of seller performance in a marketplace segmented by
service type, not price range. This is followed, in Subsection \ref{subsec:discounting-sim}, by a simulation of the effect of two discounting 
strategies on the trust measure of a generic seller. Finally, in Subsection \ref{subsec:long-term-sim} we predict, by simulation, 
the future reputation scores and trust measure of a generic seller, using currently available, incomplete information.

\subsection{Trust measure in a price-range based  multi-segment marketplace}\label{subsec:price-multi-segment}

The purpose of this subsection is to illustrate, by simulation, the trust measure of a seller in different market segments defined by the
dollar value of the goods transacted.   
For the simulation, we assume that the transactions in the marketplace are divided into four non-overlapping price ranges $R_1, R_2, R_3$, and $R_4$, 
based on the monetary value of the items transacted. These four price ranges determine four disjoint market segments—$M_1, M_2, M_3, M_4$, 
where market segment $M_i$ includes all transactions falling within the price range $R_i$.

We  have simulated a seller that has accumulated, over a time window of 250 units, the following performance in each of the four market segments:
\begin{itemize}
\item In market segment $M_1$ the seller had 85 successful transactions out of 100 total transactions;
\item In market segment $M_2$ the seller had 3 successful transactions out of 3 total transactions;
\item In market segment $M_3$ the seller had 1 successful transaction out of 1 total transaction; and,
\item In market segment $M_4$ the seller had zero transactions;
\end{itemize}

\begin{figure}[!h]
\centering
\includegraphics[clip, scale=0.44]{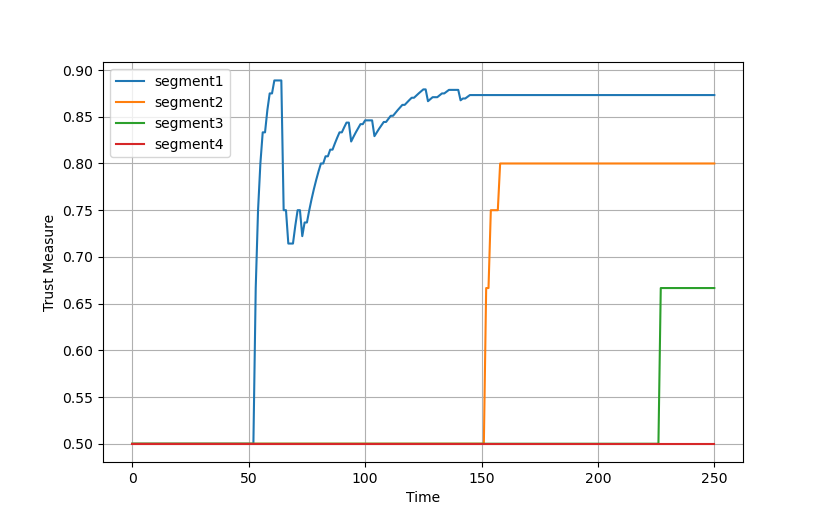}
\caption{\em Illustrating the trust measure in a price-based multi-segment market.}
\label{fig:ms-1}
\end{figure}

Figure \ref{fig:ms-1} illustrates the seller's trust measure in each of the four market segments using (\ref{eq:kn}) from  Theorem \ref{thm:kn}.

Not surprisingly, even though the trust measure of the seller in market segment $M_1$ is fairly high, 86/102, her trust measure in market 
segment $M_3$ is a meager 2/3, while in market segment $M_4$ the seller's performance is only 1/2, reflecting the fact that the seller has had 
no experience in the market segment. As a result, the seller cannot misrepresent her performance.

\subsection{Trust measure in a service-type based multi-segment marketplace}\label{subsec:service-multi-segment}

In Subsection \ref{subsec:price-range} we argued that the reputation scores and trust measure of a seller should not be global but should,
instead, be specific to individual price ranges. 
In Subsection \ref{subsec:service-type} we extended the same idea to various types of services and made the point that the best indicator of 
how the service provider will perform in the future depends on their
past performance in the context of the type of services contemplated. 

This motivated us to assess the trustworthiness of a service provider by
the type of individual service of interest. With this in mind, we have simulated the evolution of reputation scores and trust measures of a 
plumbing contractor who is offering the following services:
electric heater installation,
gas heater installation,
general plumbing repairs,
long-term maintenance contracts,
sewer repairs, and
gas boiler service.

Some of these services are more lucrative than others and our plumber is more competent in dealing with electricity than with gas equipment 
installation and repairs. Thus, our plumber may be inclined to provide higher quality services in areas that benefit him most 
(e.g. installing electric water heaters and general plumbing repairs) and of lesser quality in some other areas that are less lucrative, e.g. 
installing gas water heaters or providing sewer repairs. We note that in this case, the quality of a service is not necessarily price range
dependent, because an electric water heater may cost roughly the same as a gas water heater.

\begin{figure}[!h]
\centering
\includegraphics[clip, scale=0.44]{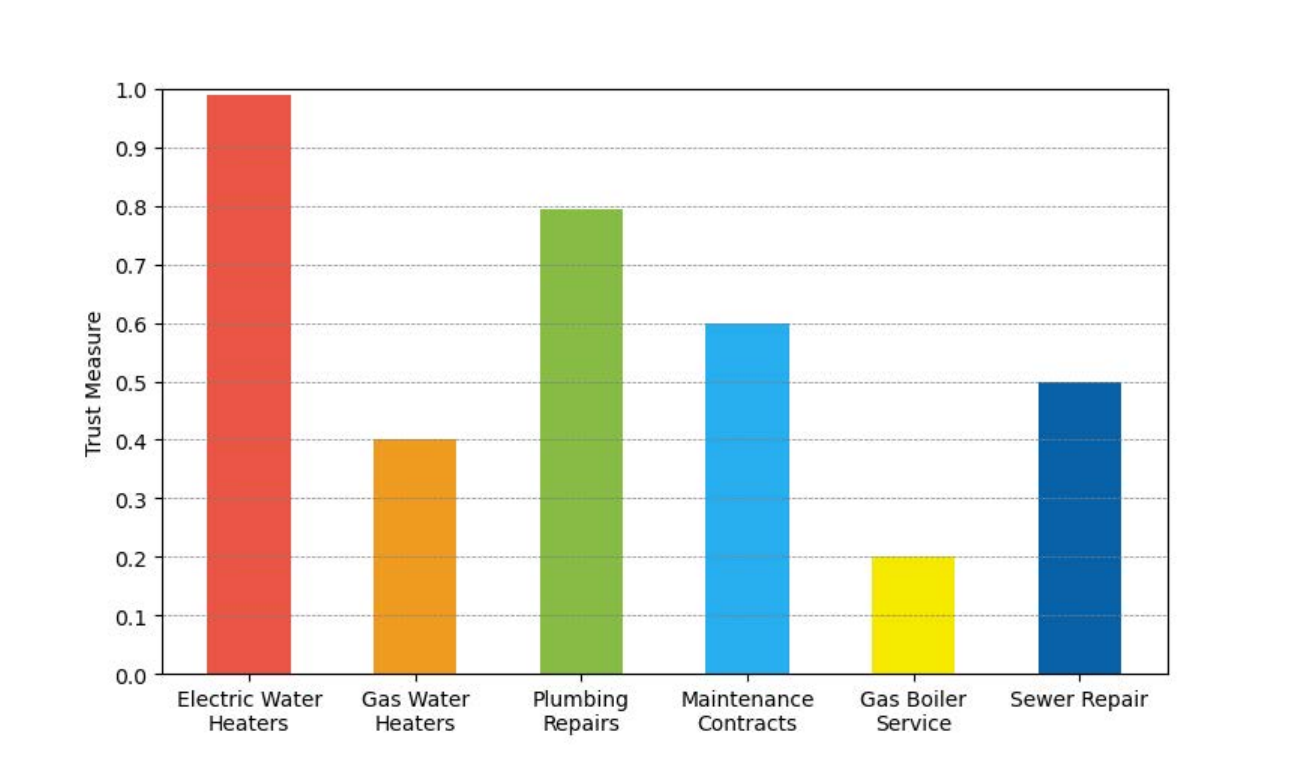}
\caption{\em Illustrating a plumber's trust measure in a service-based multi-segment market.}
\label{fig:ms-2}
\end{figure}

The point is that the plumber's  reputation score acquired by providing one type of service should not be relevant when evaluating his/her
trustworthiness in different service categories where he/she is either less competent or simply not interested in providing high-quality services.
In our simulations, our plumber has accumulated the following performance in each of the six service categories:
\begin{itemize}
\item In the electric heater installation category, the plumber  had 92 successful transactions out of 93 transactions;
\item In the gas heater installation category, the plumber  had 11  successful transactions out of 29 transactions;
\item In the general plumbing repairs category, the plumber had 39 successful transactions out of 48 transactions;
\item In the maintenance contract category, the plumber  had 58 successful transactions out of 98  transactions;
\item In the gas boiler service  category, the plumber  had 3 successful transactions out of 18 transactions;
\item In the sewer repairs category, the plumber had 0 successful transactions out of 0  transactions.
\end{itemize}
Figure \ref{fig:ms-2} illustrates our plumber's trust measure in each of the service categories above.

Thus, if a consumer wishes to hire a trustworthy plumbing contractor for gas boiler service, our plumber has nothing to recommend him
in that service category, even though they have a stellar performance in electric water heater installation.

\subsection{Illustrating the effect of discounting strategies}\label{subsec:discounting-sim}

We have simulated the reputation scores and associated trust measure, of a generic seller in six time epochs, each one week long. Each epoch has its
own success rate, as detailed below. 
\begin{itemize}
\item Epoch 1: time range from 0 to 250 with a success rate of 0.55;
\item Epoch 2: time range from 250 to 500 with a success rate of 0.65;
\item Epoch 3: time range from 500 to 750 with a success rate of 0.70;
\item Epoch 4: time range from 750 to 1000 with a success rate of 0.75;
\item Epoch 5: time range from 1000 to 1250 with a success rate of 0.80;
\item Epoch 6: time range from 1250 to 1500 with a success rate of 0.90.
\end{itemize}

Initially, the seller's reputation scores were low, perhaps because of her lack of experience. 
We have simulated the effect of the two discounting strategies presented in Section \ref{sec:extensions}. 
Table \ref{table:lambda} displays the different weights utilized as discounting parameters:

\begin{table}[h]
\centering
\begin{tabular}{l|cccccc}
 & $\lambda_1$ & $\lambda_2$ & $\lambda_3$ & $\lambda_4$ & $\lambda_5$ & $\lambda_6$ \\ \hline
Experiment 1 & 0.1 & 0.1 & 0.1 & 0.1 & 0.1 & 0.5 \\
Experiment 2 & 0.0 & 0.0 & 0.1 & 0.2 & 0.2 & 0.5 \\
Experiment 3 & 0.0 & 0.0 & 0.0 & 0.0 & 0.5 & 0.5 \\
Experiment 4 & 0.0 & 0.0 & 0.0 & 0.0 & 0.4 & 0.6 \\
Experiment 5 & 0.0 & 0.0 & 0.0 & 0.0 & 0.0 & 1.0 \\
Experiment 6 & 0.17 & 0.17 & 0.17 & 0.17 & 0.17 & 0.17 \\
\end{tabular}
\caption{Various Weights as Discounting Parameters}
\label{table:lambda}
\end{table}

\begin{figure}[!h]
\centering
\hspace*{-5mm}\includegraphics[clip, scale=0.41]{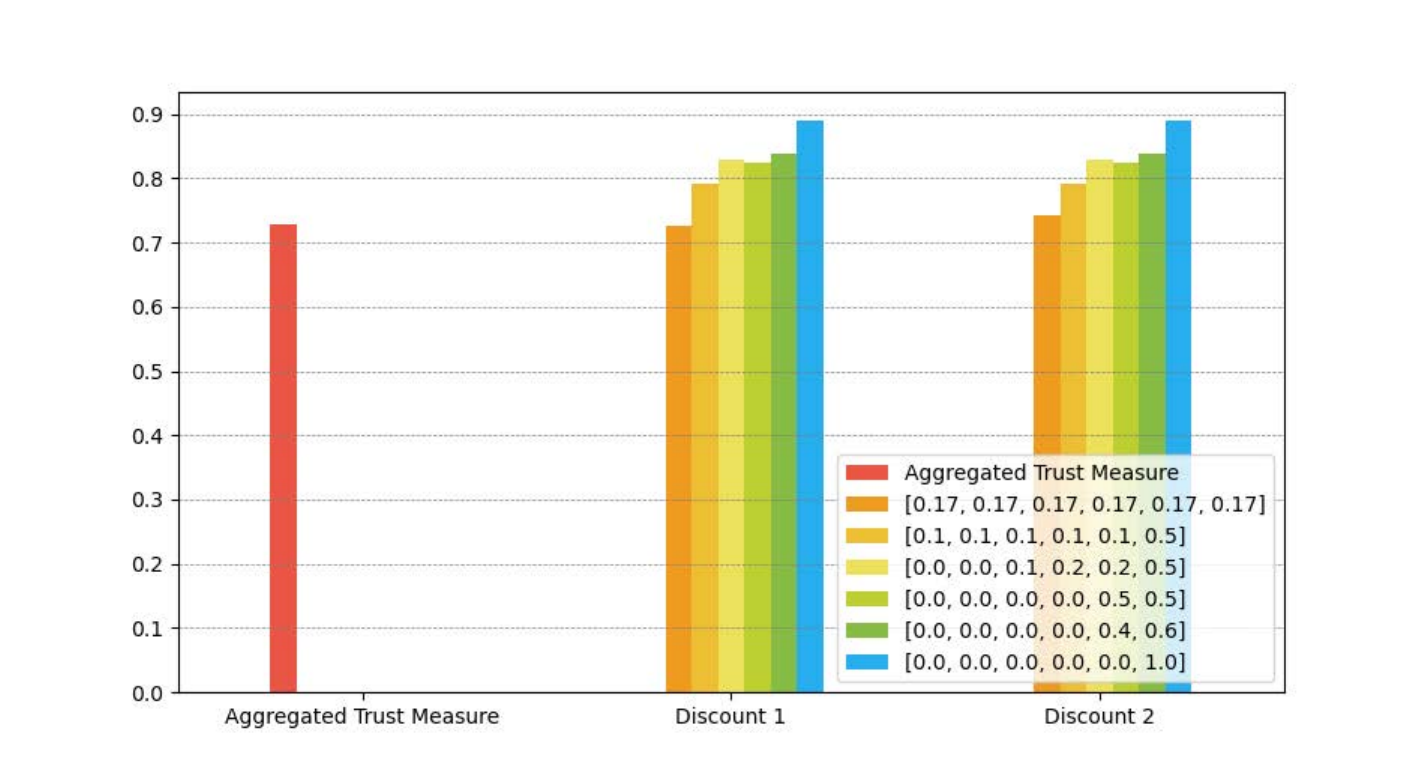}
\caption{\em A first Illustration the two discounting strategies in Subsection \ref{subsec:discount}.}
\label{fig:discount-1}
\end{figure}

The results of the simulation with each experiment color-coded are summarized
in Figure \ref{fig:discount-1}. In the figure we have plotted, side by side, the seller's aggregate trust measure without discounting as well as 
her trust measure weighted as described. In the figure, it becomes obvious the effect of favoring recent performance over more remote performance. 
As it turns out, selecting the weights that focus attention on the performance of the seller in the last week presents her trust measure in 
the best light, as it is, conceivably, the most accurate reflection of her improvement.
  
\subsection{Predicting trust measure and reputation scores over the long term}\label{subsec:long-term-sim}

\begin{figure}[!h]
\centering
\hspace*{-5mm}\includegraphics[clip, scale=0.46]{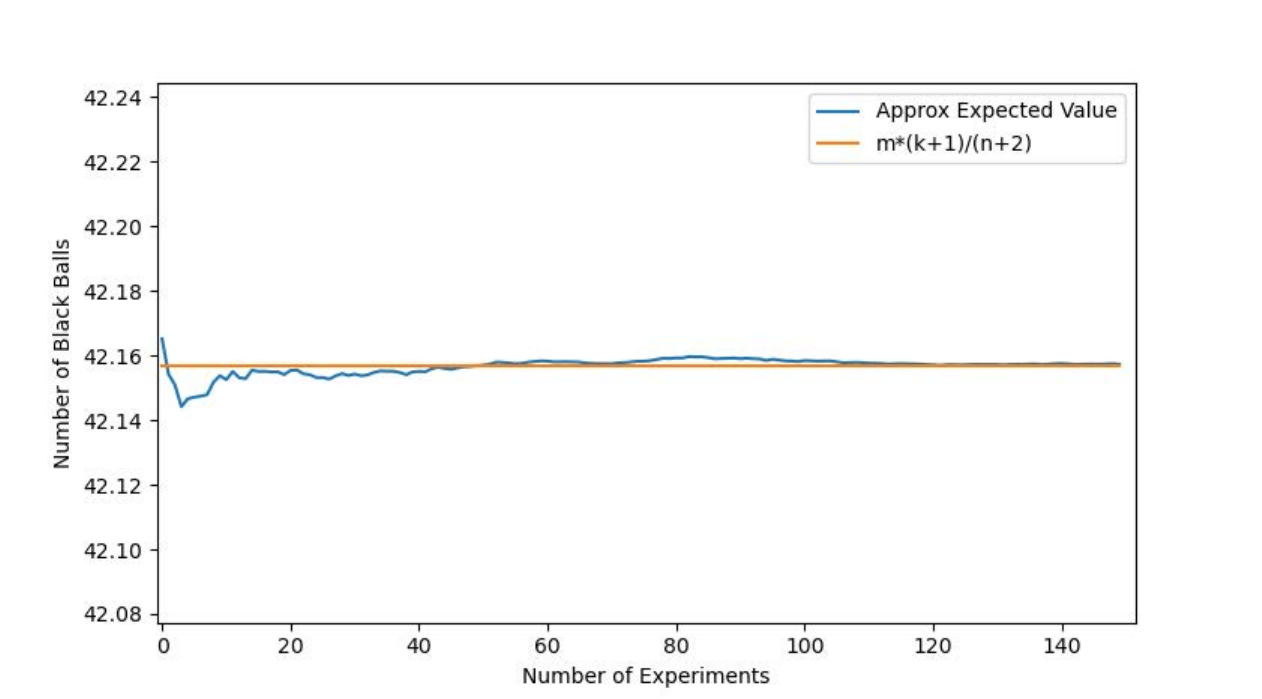}
\caption{\em Illustrating the convergence of the simulated prediction of long-term trust measure to the theoretical prediction of 
Subsection \ref{subsec:discount}.}
\label{fig:long-term}
\end{figure}

In this subsection, we are presenting the results of simulating the convergence of the predicted and simulated long-term trust measure of a seller. 
For this purpose, we have simulated the performance of a seller in her first 100 transactions. Our goal was to see how close is the prediction of 
the expected number of her successful transactions among the next 100 transactions. The results of the simulation are plotted in 
Figure \ref{fig:long-term}. The simulation was repeated between 50 and 150 times. From the figure, 
it is clear that the seller's simulated long-term performance, in terms of her reputation scores (and associated trust measure) converges to the 
theoretically predicted performance. 

\section{Concluding remarks}\label{sec:concl}
This paper was motivated by the multi-fold challenges inherent in implementing the vision of trusted and secure services in Society 5.0. 

The first main contribution of this paper was a novel trust and reputation service with a view to reducing the uncertainty associated with
buyer feedback in decentralized marketplaces. Our trust and reputation service was inspired by a classic result in probability theory that 
can be traced back to Laplace.

The second main contribution of the paper was to offer three applications of the proposed trust and
reputation service.  

Specifically, in Subsections \ref{subsec:price-range} and \ref{subsec:service-type} we discussed two applications 
to a multi-segment marketplace, where a malicious seller may establish a stellar reputation by selling cheap items or providing some specific service, 
only to use their excellent reputation score to defraud buyers in a different market segment. As we noted, our service can provide 
Sybil resistance is a much-desired attribute. 

Next, in Subsection \ref{subsec:discount}, we applied the results of
Section \ref{sec:trust-measure} in the context of sellers with time-varying performance due, for example, to fighting an initial 
learning curve or other similar impediments. We provided two discounting schemes wherein less recent reputation scores are
given less weight than more recent ones. 
In Subsection \ref{subsec:long-term} we showed how to use or trust and reputation services to predict reputation scores 
far in the future, based on fragmentary information.

Last, but certainly not least, the reputation and trust service developed in this paper seems to have applications for several 
domains, including banking, inventory management, 
vehicular networks  \cite{javaid-2019}, 
peer-to-peer networking \cite{lu-2018}, and vehicular clouds \cite{olariu-2020}.  
Exploring these new application domains promises to be an exciting area for future work.

\bibliographystyle{IEEEtran}


\vspace*{45 mm}

\begin{wrapfigure}[9]{l}[0pt]{2.4cm}
	\includegraphics[width=0.15\textwidth]{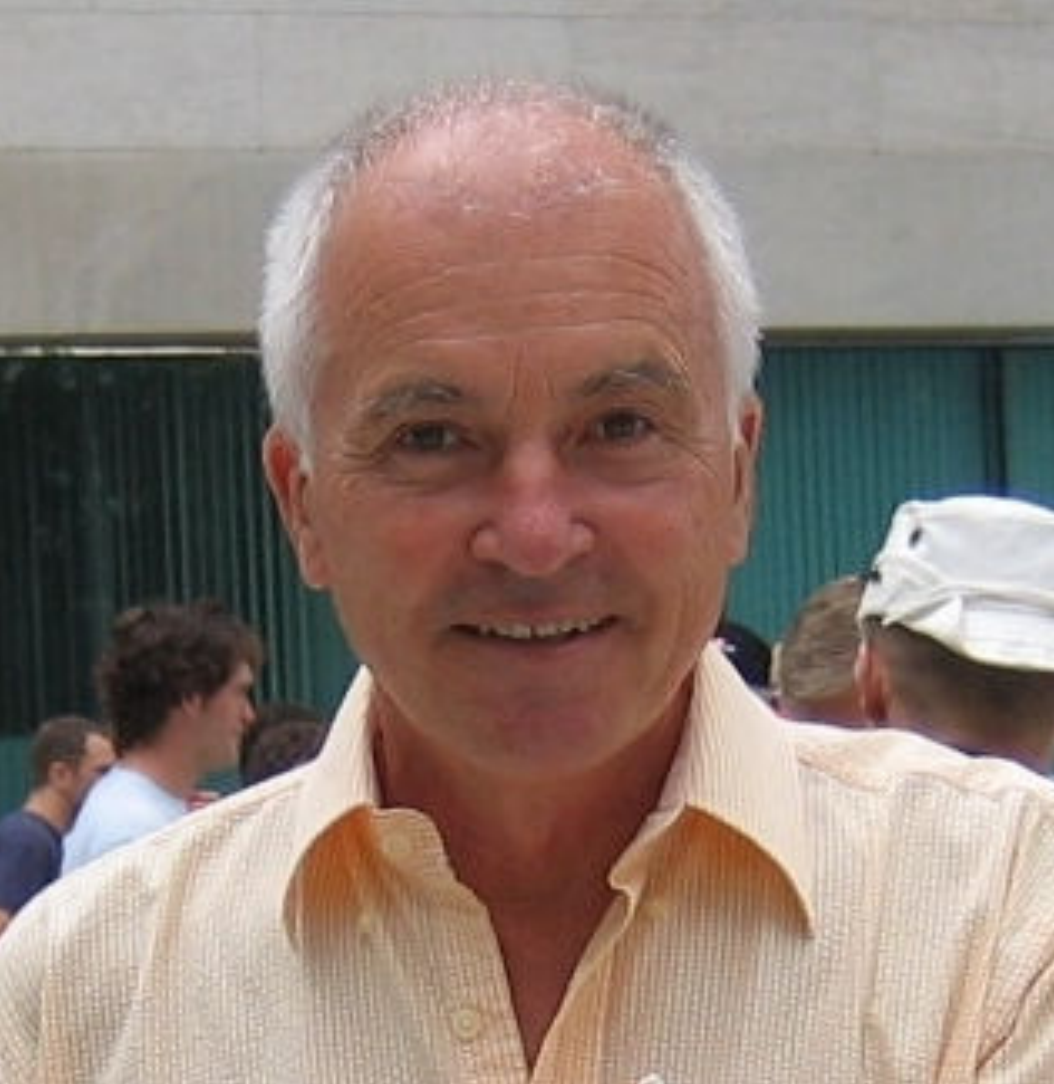}
\end{wrapfigure}
\noindent
\small{Professor Olariu received his M.Sc and PhD from McGill University, Montreal, Canada. Much of his experience has been with the 
	design and implementation of robust protocols for wireless networks and their applications.
Professor Olariu is
applying mathematical modeling and analytical frameworks to the resolution of problems ranging from securing communications
to predicting the behavior of complex systems to evaluating the performance of wireless networks. 
His most recent research interests are in the area of services computing}.
\vspace*{11mm}

\begin{wrapfigure}[9]{l}[0pt]{2.4cm}
	\vspace*{-3.6mm}
	\includegraphics[width=0.15\textwidth]{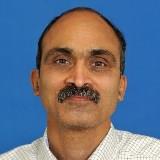}
\end{wrapfigure}
\noindent
\small{Dr. Ravi Mukkamala received his Ph.D. from the University of Iowa, Iowa City, Iowa, in 1987. 
He also received his M.B.A. from Old Dominion University in 1993. He joined ODU as an Assistant professor 
in 1987 where he is currently a Professor of Computer Science and Associate Dean for the College of Sciences.  
Dr. Mukkamala's current areas of research include computer security, privacy, data mining, and modeling. He has published 
more than 175 research papers in refereed journals and conference proceedings. He has received more than \$3 million 
in research grants as PI or co-PI from agencies including NASA, as well as Jefferson Lab and private industries. 
He received a Most Inspirational Faculty Award from ODU in 1994. He has won several best paper awards at national and 
international conferences over the years.}
\vspace*{11mm}

\begin{wrapfigure}[9]{l}[0pt]{2.4cm}
	\vspace*{-3.0mm}
	\includegraphics[width=0.15\textwidth]{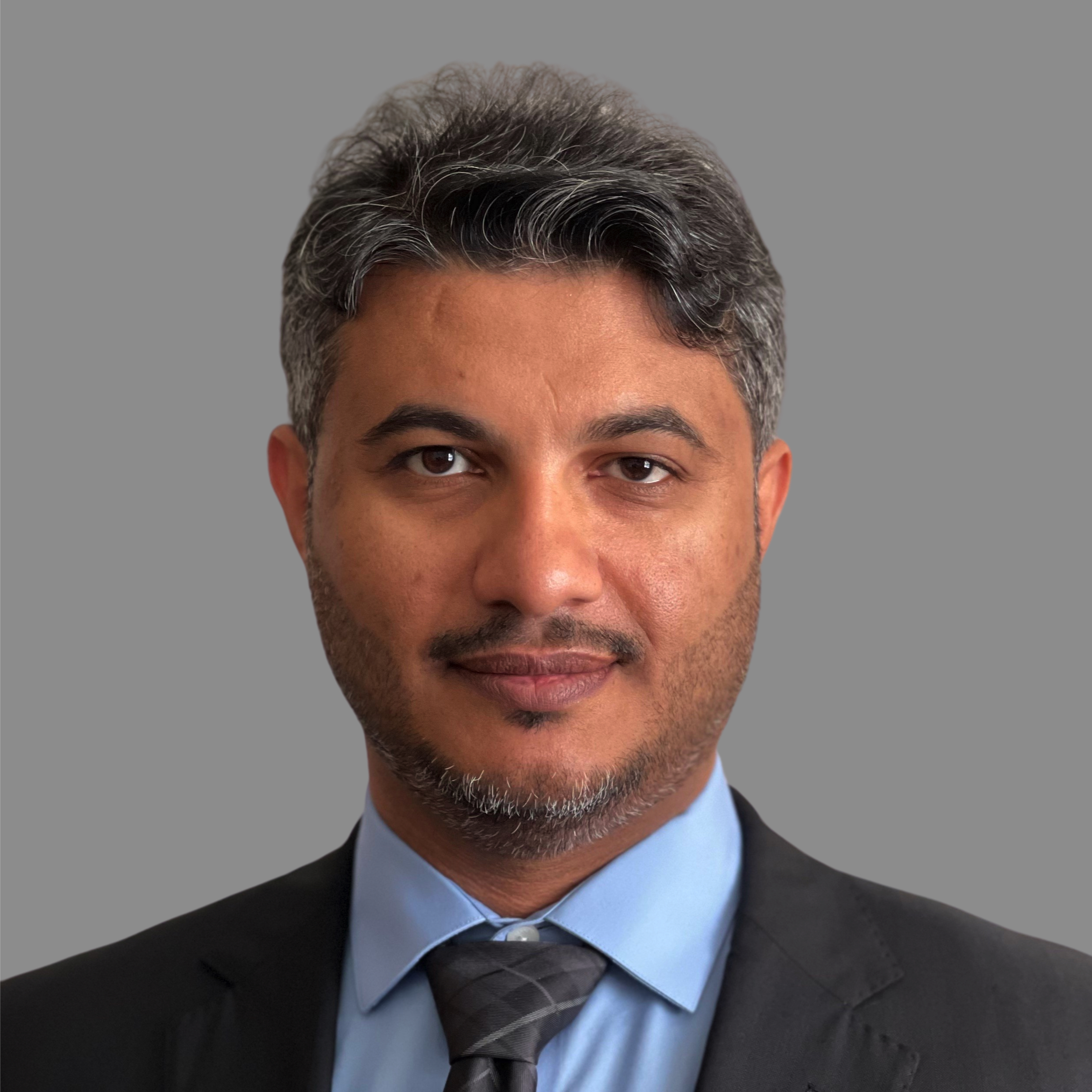}
\end{wrapfigure}
\noindent
\small{Meshari Aljohani received the M.S. degree in computer science from California Lutheran University in 2013. He is currently pursuing the Ph.D. 
       degree in computer science at Old Dominion University. His research interests are related to blockchain, marketplace, and reputation 
systems.}

\clearpage

\appendix

\setcounter{equation}{20}


\section{Appendices}\label{sec:appendices}

\subsection{Combinatorial preliminaries}\label{subsec:stepping-stones}

%
%

\begin{lemma}\label{common-term}
For non-negative integers, $r,\ s,\ t$, the following holds
\begin{equation*}
{r \choose s} {{r-s} \choose t} = {{s+t} \choose s} {r \choose {s+t}}.
\end{equation*}
\end{lemma}

\begin{IEEEproof}
See \cite{graham-1994}, pp. 167--8.
\end{IEEEproof}

\begin{lemma}\label{lovasz-lemma}
For all non-negative integers $k,\ r,\ s,\ m,\ n$, with $0 \leq r \leq n$, the following equality holds
\begin{equation}\label{eq:lovasz-1}
\sum_{k=0}^s {{r+k} \choose n} { {s-k} \choose m} = { {r+s+1} \choose {n + m +1}}.
\end{equation}
\end{lemma}

\begin{IEEEproof} 
See \cite{graham-1994}, p. 169.
\end{IEEEproof}

\subsection{Evaluating $\sum_{j=0}^N \Pr[A' | H_j]\Pr[H_j]$}\label{eval-A'}

To simplify notation, we write $\Pr[H_i]$ instead of $\Pr[H_i|n,k]$. Obviously,
Recall that by (\ref{eq:update-11}), $\Pr[H_i] =\frac { {i \choose {k}} {{N-i} \choose {n-k}}} {{{N+1} \choose {n+1}}}$ and that
$\Pr[A' | H_i] = \frac { {{i-k} \choose {k'}} {{N-i-(n-k)} \choose {n'-k'}}} {{{N-n} \choose n'}}$.
With this, the expression of $\sum_{j=0}^N \Pr[A' | H_j]\Pr[H_j]$ becomes:
\begin{eqnarray}\label{A'-1}
\sum_{j=0}^N \Pr[A' | H_j]\Pr[H_j]\hspace*{-3mm}  &=& \hspace*{-3mm}\frac{\sum_{i=0}^N {{i \choose k}} {{N-i} \choose {n-k}} {{i-k} \choose {k'}} {{{N-i- (n-k}} \choose {n'-k'}}} 
             {{{N-n} \choose {n'}} {{N+1} \choose {n+1}}}  
\end{eqnarray}

\noindent
By Lemma \ref{common-term} in the Appendix \ref{subsec:stepping-stones} we can write
\begin{equation}\label{s1}
{i \choose k} {{i-k} \choose {k'}} = {{k+k'} \choose k} { i \choose {k+k'}}
\end{equation}
and
\begin{equation}\label{s2}
\scriptstyle {{N-i} \choose {n-k}} {{N-i -(n-k)} \choose {n'-k'}} = {{n-k + n'-k'} \choose {n-k}} { {N-i} \choose {n-k+n'-k'}}.
\end{equation}

On replacing (\ref{s1}) and (\ref{s2}) back into (\ref{A'-1}) we obtain
\begin{eqnarray}\label{A'-2}
\sum_{j=0}^N \Pr[A' | H_j]\Pr[H_j]\hspace*{-3mm} &=&\hspace*{-3mm} \frac{{{k+k'} \choose k} {{n-k + n'-k'} \choose {n-k}}} 
        {{{N-n} \choose {n'}} {{N+1} \choose {n+1}}} \scriptstyle \sum_{i=0}^N { i \choose {k+k'}} { {N-i} \choose {n-k+n'-k'}} \nonumber \\
        &=& \frac{{{k+k'} \choose k} {{n-k + n'-k'} \choose {n-k}} {{N+1} \choose {n+n'+1}}}
        {{{N-n} \choose {n'}} {{N+1} \choose {n+1}}} \nonumber \\
        &=& \frac{ {{k+k'} \choose k} {{n-k + n'-k'} \choose {n-k}} } {{{n+n'+1} \choose {n+1}}}.
\end{eqnarray} 

\subsection{Two simple algebraic inequalities}\label{subsec:alg}

\begin{lemma}\label{lem:a0b0}
Let $r$ be an arbitrary positive integer and consider non-negative real numbers $a_0, a_1, \cdots, a_r$ and $\alpha_1, \alpha_2, \cdots, \alpha_r$ as well as
positive reals $b_0, b_1, \cdots, b_r$ such that
\begin{equation}\label{eq:a0b0}
\frac{a_i}{b_i} \leq \frac{a_0}{b_0},\ {\rm for\ all}\ i=1,\ 2,\ \cdots,\ r.
\end{equation}
Then,
\begin{equation}\label{eq-a0b0-1}
\frac{\sum_{i=1}^r \alpha_i~a_i}{\sum_{i=1}^r \alpha_i~b_i} \leq \frac{a_0}{b_0}.
\end{equation}
\end{lemma}

\begin{IEEEproof}
Let us evaluate the difference
\begin{eqnarray}\label{eqn:a0b0}
\frac{a_0}{b_0} - \frac{\sum_{i=1}^r \alpha_i~a_i}{\sum_{i=1}^r \alpha_i~b_i} &=& \frac{a_0 \sum_{i=1}^r \alpha_i~b_i - b_0 \sum_{i=1}^r \alpha_i~a_i}
 {b_0 \sum_{i=1}^r \alpha_i~b_i}\nonumber \\
      &=& \frac{\sum_{i=1}^r \alpha_i~\left [ a_0~b_i - b_0~a_i \right ]}
 {b_0 \sum_{i=1}^r \alpha_i~b_i}\nonumber \\
      &=& \frac{\sum_{i=1}^r \alpha_i~b_0~b_i \left [ \frac{a_0}{b_0} - \frac{a_i}{b_i} \right ]}
 {b_0 \sum_{i=1}^r \alpha_i~b_i}\nonumber \\
      &=& \frac{\sum_{i=1}^r \alpha_i~b_i \left [ \frac{a_0}{b_0} - \frac{a_i}{b_i} \right ]}
 {\sum_{i=1}^r \alpha_i~b_i}\nonumber \\
      & \geq& 0\ \ \ \ \ \mbox{[by (\ref{eq:a0b0}).]}
\end{eqnarray}

\end{IEEEproof}
\begin{lemma}\label{lem:aa'bb'}
Let $a,\ a'$ be non-negative reals and let $b,\ b'$ be positive reals. Then
\begin{equation}\label{eq:aa'bb'}
\frac{a +a'}{b+b'} \geq \frac{a}{b} \iff \frac{a'}{b'} \geq \frac{a}{b}.
\end{equation}
\end{lemma}
\begin{IEEEproof}
We write in stages
\begin{eqnarray*}
\frac{a +a'}{b+b'} \geq \frac{a}{b} &\iff& \frac{a+a'}{a} \geq \frac{b+b'}{b} \nonumber \\
                                    &\iff& \frac{a'}{a} \geq \frac{b'}{b} \nonumber \\
                                    &\iff& \frac{a'}{b'} \geq \frac{a}{b},
\end{eqnarray*}
and the proof of the lemma is complete.
\end{IEEEproof}

\end{document}